\documentclass[prl,aps,showpacs,twocolumn,superscriptaddress,nofootinbib]{revtex4}
\usepackage{mathrsfs}
\usepackage{amssymb}
\usepackage{amsmath}
\usepackage{graphicx}
\usepackage[caption = false]{subfig}
\usepackage[normalem]{ulem}
\usepackage{xcolor}
\usepackage[breaklinks,colorlinks,urlcolor=blue,citecolor=citcolor,linkcolor=lcolor]{hyperref}
\usepackage[utf8]{inputenc}
\usepackage{slashed}
\usepackage{tikz}

\definecolor{lcolor}{rgb}{0.5,0,0}
\definecolor{citcolor}{rgb}{0,0.3,0.0}

\newcommand{\n}{\nonumber \\}

\makeindex

\begin{document}

\title{Diffusion of conserved charges in relativistic heavy ion collisions}
\author{Moritz Greif}
\affiliation{Institut f\"ur Theoretische Physik, Johann Wolfgang Goethe-Universit\"at,
Max-von-Laue-Str.\ 1, D-60438 Frankfurt am Main, Germany}
\email{greif@th.physik.uni-frankfurt.de}
\author{Jan. A. Fotakis}
\affiliation{Institut f\"ur Theoretische Physik, Johann Wolfgang Goethe-Universit\"at,
Max-von-Laue-Str.\ 1, D-60438 Frankfurt am Main, Germany}
\author{Gabriel S. Denicol}
\affiliation{Instituto de F\'{\i}sica, Universidade Federal Fluminense, UFF, Niter\'{o}i,
24210-346, RJ, Brazil}
\author{Carsten Greiner}
\affiliation{Institut f\"ur Theoretische Physik, Johann Wolfgang Goethe-Universit\"at,
Max-von-Laue-Str.\ 1, D-60438 Frankfurt am Main, Germany}
\date{\today }

\begin{abstract}
In order to characterize nuclear matter under extreme conditions, we calculate all diffusion transport coefficients related to baryon,
electric and strangeness charge for both a hadron resonance gas and a simplified kinetic model of the quark-gluon plasma.
We demonstrate that the diffusion currents do not depend only on gradients of their
corresponding charge density. Instead, we show that there exists coupling
between the different charge currents, in such a way that it is possible for
density gradients of a given charge to generate dissipative currents of
another charge. Within this scheme, the charge diffusion coefficient is best
viewed as a matrix, in which the diagonal terms correspond to the usual
charge diffusion coefficients, while the off-diagonal terms describe the
coupling between the different currents. In this letter, we calculate for
the first time the complete diffusion matrix including the three charges
listed above. 
We find that the baryon diffusion current is strongly affected by baryon charge gradients, but also by its coupling to gradients in strangeness. The electric charge diffusion current is found to be strongly affected by electric and strangeness gradients, whereas strangeness currents depend mostly on strange and baryon gradients.
\end{abstract}

\maketitle



\paragraph*{Introduction.}

\label{sec:Intro} Ultrarelativistic hadronic collisions, performed in the
largest particle accelerators in the world, create conditions required
to study the properties of hot and dense hadronic and quark matter. In the
last 10 years, these experiments have played a crucial role in uncovering novel transport properties of the quark-gluon plasma (QGP), a new state
of nuclear matter in which quarks and gluons are no longer confined inside
hadrons. In particular, several phenomenological studies \cite{Romatschke:2007mq,Xu:2007jv,Luzum:2008cw,Bozek:2009dw,Song:2010mg, Niemi:2011ix,Wesp:2011yy,Gale:2012rq} demonstrated that the QGP
has one of the smallest shear viscosity to entropy density ratios in nature
-- a surprising result that is still not well understood from first
principles. Additional theoretical and phenomenological studies \cite{Arnold:2006fz,Kharzeev:2007wb,Bozek:2009dw,Noronha-Hostler:2013gga,Nopoush:2014pfa,Denicol:2014vaa,Ryu:2015vwa}
have also improved our understanding of the bulk viscosity, showing that
this coefficient can display novel behavior near the deconfinement
transition of nuclear matter. 
Recently, much attention was paid to the electric conductivity;
several studies on the lattice \cite{Aarts:2014nba,Brandt:2015aqk,Ding:2016hua}, in
perturbative QCD (pQCD) \cite{Greif:2014oia,Puglisi:2014sha,Arnold:2000dr} and
effective theories \cite{Rougemont:2015ona,Greif:2016skc,Rougemont:2017tlu} have been carried out.

On the other hand, at this stage very little is known about net-charge
diffusion in hot and dense nuclear matter. This is due to the fact that in high
energy heavy ion collisions, the net-charge density of the matter produced
is extremely small in almost all space-time points and it becomes very
difficult to observe any dissipative effects due to diffusion \cite{Monnai:2012jc}. Recently, the Relativistic Heavy-Ion Collider (RHIC)
started to perform hadronic collisions at lower energies within the beam energy
scan (BES) program in order to investigate the phase diagram and transport properties of
nuclear matter at finite net-baryon (and net-electric charge) density~\cite{Aggarwal:2010cw,Mohanty:2011nm,Mitchell:2012mx}. The central plateau in the rapidity distribution of baryon multiplicity $\mathrm{d}N_B/d\eta$ is less pronounced at those low energies, such that strong gradients in the chemical potential of conserved charges are expected.
At beam energies down to, e.g., $\sqrt{s_{\mathrm{NN}}}=7.7~\mathrm{GeV}$ in the RHIC BES, the baryon chemical potential can reach values up to $\mu_B\sim 400~\mathrm{MeV}$ which is significant compared to the temperatures that are reached~\cite{Odyniec:2013kna, Adamczyk:2017iwn}. Therefore, one can expect that low energy collisions can be particularly useful to explore the properties of net-charge diffusion of nuclear matter, that were out of reach in higher
energy collisions.

In the relativistic Navier-Stokes-Fourier theory, a net-charge ($q$)
diffusion 4-current, $j_{q}^{\mu }$, is determined by the following
constitutive relation, 
\begin{equation}
j_{q}^{\mu }=\kappa _{q}\nabla ^{\mu }\alpha _{q},
\end{equation}%
where $\alpha _{q}\equiv \mu _{q}/T$ is the thermal potential, with $\mu _{q}
$ being the charge chemical potential, $T$ the temperature and $\kappa_q$ the corresponding net-charge diffusion coefficient.
We further
defined the transverse gradient $\nabla ^{\mu }\equiv \Delta ^{\mu \nu
}\partial _{\nu }$, and the projection operator $\Delta ^{\mu \nu }\equiv
g^{\mu \nu }-u^{\mu }u^{\nu }$, where $u^{\mu }$ is the local fluid velocity
and  $g^{\mu \nu }$ the space-time metric. We remark that this relativistic
constitutive relation does not only describe the effects of charge diffusion but
also includes the effects of heat flow. 

It is also important to emphasize that the constitutive relations satisfied
by the diffusion 4-currents become different in the presence of more than
one conserved charge. When discussing charge diffusion in the matter
produced in heavy ion collisions, we must consider at least three conserved
charges: baryon number (B), electric charge (Q), and strangeness (S). Since
several hadrons (and quarks) carry more than one of these charges, the
diffusion current of each charge will no longer be solely proportional to the
gradient of the thermal potential ($\nabla ^{\mu }\alpha _{q}$) of that specific charge. Instead, there
will be a mixing between the currents, with gradients of every single charge
density being able to generate a diffusion current of any other charge. In
general, one has 
\begin{equation}
\begin{pmatrix}
\begin{tabular}{c}
$j_{B}^{\mu }$ \\ 
$j_{Q}^{\mu }$ \\ 
$j_{S}^{\mu }$%
\end{tabular}%
\end{pmatrix}%
=%
\begin{pmatrix}
\begin{tabular}{ccc}
$\kappa _{BB}$ & $\kappa _{BQ}$ & $\kappa _{BS}$ \\ 
$\kappa _{QB}$ & $\kappa _{QQ}$ & $\kappa _{QS}$ \\ 
$\kappa _{SB}$ & $\kappa _{SQ}$ & $\kappa _{SS}$%
\end{tabular}%
\end{pmatrix}%
\cdot 
\begin{pmatrix}
\begin{tabular}{c}
$\nabla ^{\mu }\alpha _{B}$ \\ 
$\nabla ^{\mu }\alpha _{Q}$ \\ 
$\nabla ^{\mu }\alpha _{S}$%
\end{tabular}%
\end{pmatrix}%
,  \label{eq:LinearDiff}
\end{equation}%
leading to a diffusion coefficient that is a matrix instead of a number, $%
\kappa _{qq^{\prime }}$. Therefore, it is not sufficient to compute what is
usually known as baryon, electric and strangeness diffusion coefficients,
i.e., the diagonal terms in the matrix $\kappa _{BB},\kappa _{QQ},\kappa
_{SS}$, but also, one must calculate the off-diagonal terms or couplings
terms $\kappa _{QB},\kappa _{SB},\kappa _{SQ}$ (it is sufficient to
calculate only these three off-diagonal terms, since Onsager's theorem~\cite%
{PhysRev.37.405,PhysRev.38.2265} guarantees that the diffusion matrix is
symmetric).

The dynamics of the thermal potentials $\alpha_B,\alpha_S$ and $\alpha_Q$ and their respective currents in heavy ion collisions is currently not well known. It is expect that the influence of diffusion currents on the hydrodynamical evolution of the net-charge currents can be very pronounced at lower collision energies, leading to significant effects on certain observables ~\cite{Monnai:2012jc}. In this letter, we calculate for the first time the complete charge
diffusion matrix, for the three charges listed above. We perform this
task for a hadron resonance gas (HRG) and for a kinetic theory toy model of
the QGP. We find that the coupling terms can be as large as
the diagonal terms and, consequently, models simulating heavy ion
collisions including only the diagonal contributions to net-charge diffusion
may be missing crucial ingredients. Furthermore, it may not be a good approximation to perform simulations
including only the dynamics of one charge since its gradients will necessarily give rise to diffusion
currents of the remaining charges. We use natural units, $\hbar =c=k_{B}=1$ and
Minkowski metric $g^{\mu \nu }=(1,-1,-1,-1)$. Greek indices run from $0$ to $%
4$.

\paragraph*{First order Chapman-Enskog expansion.}

\label{Chapman} We consider a dilute gas consisting of $N_{\text{species}}$
particle species (either hadrons or quarks and gluons), with the $i$-th
particle species having degeneracy $g_{i}$, electric charge $Q_{i}$,
strangeness charge $S_{i}$, baryonic charge $B_{i}$ and 4-momentum $%
k_{i}^{\mu }$. The state of the system is characterized by the
single-particle momentum distribution function of each particle species, $%
f_{i}(x,k)\equiv f_{\mathbf{k}}^{i}$, with the time evolution of $f_{\mathbf{%
k}}^{i}$ being given by the relativistic Boltzmann equation. In contrast to
previous work \cite{Greif:2016skc}, we disregard any external field.

The single-particle distribution of each particle species is expanded in a
Chapman-Enskog series, i.e., in a gradient expansion~\cite{chapman1970mathematical,Denicol:2014loa}. In this case, the
Boltzmann equation is written as%
\begin{equation}
\epsilon k_{i}^{\mu }\partial _{\mu }f_{\mathbf{k}}^{i}=-\sum%
\limits_{j=1}^{N_{\text{species}}}C_{ij}(x^{\mu },k^{\mu }),
\end{equation}%
with $C_{ij}(x^{\mu },k^{\mu })$ being the collision term and $\epsilon $ a
book-keeping parameter that will be set to one at the end of the
calculation. The Chapman-Enskog expansion is just an expansion in powers of $%
\epsilon $, $f_{\mathbf{k}}^{i}\sim f_{0\mathbf{k}}^{i}+\epsilon f_{1\mathbf{%
k}}^{i}+\epsilon ^{2}f_{2\mathbf{k}}^{i}+\cdots $, where $f_{j\mathbf{k}%
}^{i} $ is the $j$--th order solution of the expansion. The zeroth order
solution of this series is the local equilibrium distribution function, leading to the equations of ideal fluid dynamics, while the first order solution contains terms that are of first order in
gradients of velocity, temperature and chemical potential, leading to the
equations of relativistic Navier-Stokes theory and the diffusion equation 
\cite{chapman1970mathematical,Denicol:2014loa}. For the purposes of this letter, it is
sufficient to calculate the first order contribution, which is the order
that determines the diffusion coefficients. Without loss of generality, we
only retain the terms of the expansion that contribute directly to the
diffusion terms, omitting all others that contribute to shear and bulk
viscosity.

The equation for the first-order Chapman-Enskog correction will then be 
\begin{equation}
-\sum\limits_{j=1}^{N_{\text{species}}}\hat{C}_{ij}^{(1)}f_{1\mathbf{k}%
}=\sum_{q\in \{B,Q,S\}}f_{0\mathbf{k}}^{i}k_{i}^{\mu }\nabla _{\mu }\alpha
_{q}\left( \frac{E_{i,\mathbf{k}}n_{q}}{\epsilon _{0}+P_{0}}-q_{i}\right) .\
\,  \label{eq:Boltzmann}
\end{equation}%
where $\hat{C}_{ij}^{(1)}$ is the linearized collision operator, 
\begin{eqnarray}
\hat{C}_{ij}^{(1)}f_{1\mathbf{k}} &=&\int dK^{\prime }dPdP^{\prime }\gamma
_{ij}W_{\mathbf{kk}^{\prime }-\mathbf{pp}^{\prime }}^{ij}f_{0\mathbf{k}%
}^{i}f_{0\mathbf{k}^{\prime }}^{j}  \label{collisionterm} \\
&&\times \left( \frac{f_{1\mathbf{p}}^{i}}{f_{0\mathbf{p}}^{i}}+\frac{f_{1%
\mathbf{p}^{\prime }}^{j}}{f_{0\mathbf{p}^{\prime }}^{j}}-\frac{f_{1\mathbf{k%
}}^{i}}{f_{0\mathbf{k}}^{i}}-\frac{f_{1\mathbf{k}^{\prime }}^{j}}{f_{0%
\mathbf{k}^{\prime }}^{j}}\right) ,
\end{eqnarray}%
with $W_{\mathbf{kk^{\prime }}\rightarrow \mathbf{pp^{\prime }}}^{ij}=(2\pi
)^{6}s~\sigma ^{ij}(s,\theta )\delta ^{(4)}(p_{i}+p_{j}^{\prime
}-k_{i}-k_{j}^{\prime })$ being the scattering amplitude, $\sigma
^{ij}(s,\theta )$ the differential cross section, and $\gamma _{ij}=1-\delta
_{ij}/2$. For the sake of simplicity, we only consider elastic, $%
2\leftrightarrow 2$, collisions between hadrons or quarks and employ
classical statistics.

This equation can be solved following the well known procedure outlined in 
\cite{Denicol:2011fa,Greif:2016skc}. Since the collision operator, $\hat{C}%
_{ij}^{(1)}$, is linear, the solution for $f_{1\mathbf{k}}$ must be of the
general form $f_{1\mathbf{k}}^{i}=\sum_{q}a_{q}^{i}k_{i}^{\mu }\nabla _{\mu
}\alpha _{q}$, where the coefficient $a_{q}^{i}$ is a function of the energy
in the local rest frame, $E_{i,\mathbf{k}}\equiv u_{\mu }k_{i}^{\mu }$.
Next, one expands $a_{q}^{i}$ in powers of energy, $f_{1\mathbf{k}%
}^{i}=\sum_{q}k_{i}^{\mu }\nabla _{\mu }\alpha _{q}\sum_{m=0}^{M}a_{q,m}^{i}$
$\left( E_{i,\mathbf{k}}\right) ^{m}$, where the integer $M$ characterizes
the truncation of the Taylor series. Finally, one substitutes this
expansion into Eq.~\eqref{eq:Boltzmann}, multiplies the equation by the $n$%
--th basis element of the expansion, $\Delta _{\nu }^{\mu }k_{i}^{\nu
}\left( E_{i,\mathbf{k}}\right) ^{n}$, and integrates it in momentum space.
This leads to the following equation for the expansion coefficients, $%
a_{q,m}^{j}$, 
\begin{equation}
\sum_{m=0}^{M}\sum_{j=1}^{N_{\text{Species}}}\left( A_{nm}^{i}\delta
^{ij}+C_{nm}^{ij}\right) a_{q,m}^{j}=b_{q,n}^{i}\ \ ,
\label{eq:LinearSystem}
\end{equation}
where we defined 
\begin{widetext}
\begin{align}
b^i_{q,n} &= \int \frac{\mathrm{d}^3 k_i}{(2\pi)^3 E_{i,\mathbf{k}}} f_{0\mathbf{k}}^{i}  \left(\frac{E_{i,\mathbf{k}} n_q}{\epsilon_0 + P_0} - q_i\right) E^{n-1}_{i,\mathbf{k}} \Delta_{\mu\nu} k_i^\mu k_i^\nu,  \n
\mathcal{A}^i_{nm} &= \sum_{j=1}^{N_{\text{Species}}} \int \frac{\mathrm{d}^3 k_i}{(2\pi)^3 E_{i,\mathbf{k}}} \frac{\mathrm{d}^3 k^{\prime}_j}{(2\pi)^3 E_{j,\mathbf{k^\prime}}} \frac{\mathrm{d}^3 p_i}{(2\pi)^3 E_{i,\mathbf{p}}} \frac{\mathrm{d}^3 p^{\prime}_j}{(2\pi)^3 E_{j,\mathbf{p^\prime}}} \gamma_{ij} W^{ij}_{\mathbf{kk^\prime}\rightarrow\mathbf{pp^\prime}} f_{0\mathbf{k}}^{i} f_{0\mathbf{k}^\prime}^{j} E_{i,\mathbf{k}}^{n-1} \Delta_{\mu\nu}  k_i^{\mu} \left(E^m_{i,\mathbf{p}} p_i^{\nu} - E^m_{i,\mathbf{k}} k_i^{\nu}   \right), \label{eq:Matrices} \n
\mathcal{C}^{ij}_{nm} &= \int \frac{\mathrm{d}^3 k_i}{(2\pi)^3 E_{i,\mathbf{k}}} \frac{\mathrm{d}^3 k^{\prime}_j}{(2\pi)^3 E_{j,\mathbf{k^\prime}}} \frac{\mathrm{d}^3 p_i}{(2\pi)^3 E_{i,\mathbf{p}}} \frac{\mathrm{d}^3 p^{\prime}_j}{(2\pi)^3 E_{j,\mathbf{p^\prime}}} \gamma_{ij} W^{ij}_{\mathbf{kk^\prime}\rightarrow\mathbf{pp^\prime}} f_{0\mathbf{k}}^{i} f_{0\mathbf{k}^\prime}^{j} E_{i,\mathbf{k}}^{n-1} \Delta_{\mu\nu}  k_i^{\mu} \left(E^m_{j,\mathbf{p^\prime}} {p^\prime_j}^{\nu} - E^m_{j,\mathbf{k^\prime}} {k^\prime_j}^{\nu}   \right).
\end{align}
\end{widetext}
In this work, the expansion in powers of energy is truncated
at the lowest level possible, by setting $M=1$. This assumption is mainly
employed to simplify the numerical calculations we perform. Nevertheless, we
have checked, in simpler examples solved using constant cross sections, that
higher truncation values lead to only small corrections to the diffusion
coefficients, as was also demonstrated in previous work \cite
{Denicol:2011fa,Greif:2016skc} for other transport coefficients.

The $q$--th charge diffusion current is given as 
\begin{equation}
j_{q}^{\mu }=\sum_{i=1}^{N_{\text{Species}}}q_{i}\int \frac{\mathrm{d}%
^{3}k_{i}}{(2\pi )^{3}E_{i,\mathbf{k}}}\Delta _{\nu }^{\mu }k_{i}^{\nu }f_{1%
\mathbf{k}}^{i}.  \label{eq:KineticDiff}
\end{equation}%
Substituting the expansion for $f_{1\mathbf{k}}^{i}$ into Eq.~%
\eqref{eq:KineticDiff}, and comparing to Eq.~\eqref{eq:LinearDiff}, leads to
the following expression for the diffusion coefficients 
\begin{align}
\kappa _{qq^{\prime }}=\frac{1}{3}\sum_{i=1}^{N_{\text{Species}}}&
q_{i}\sum_{m=0}^{M}a_{q^{\prime },m}^{i}  \notag \\
& \times \int \frac{\mathrm{d}^{3}k_{i}}{(2\pi )^{3}E_{i,\mathbf{k}}}E_{i,%
\mathbf{k}}^{m}\Delta _{\mu \nu }k_{i}^{\mu }k_{i}^{\nu }f_{0\mathbf{k}}^{i}.
\label{eq:DiffMatrix}
\end{align}%
Therefore, calculating $\kappa _{qq^{\prime }}$ is reduced to evaluating the
integrals in Eq.~\eqref{eq:Matrices} and then solving the set of linear
equations satisfied by $a_{q^{\prime },m}^{i}$ in Eq.~\eqref{eq:LinearSystem}%
. Both these tasks are performed numerically.

\begin{figure}[tbp]
\centering
\includegraphics[width=0.5\textwidth]{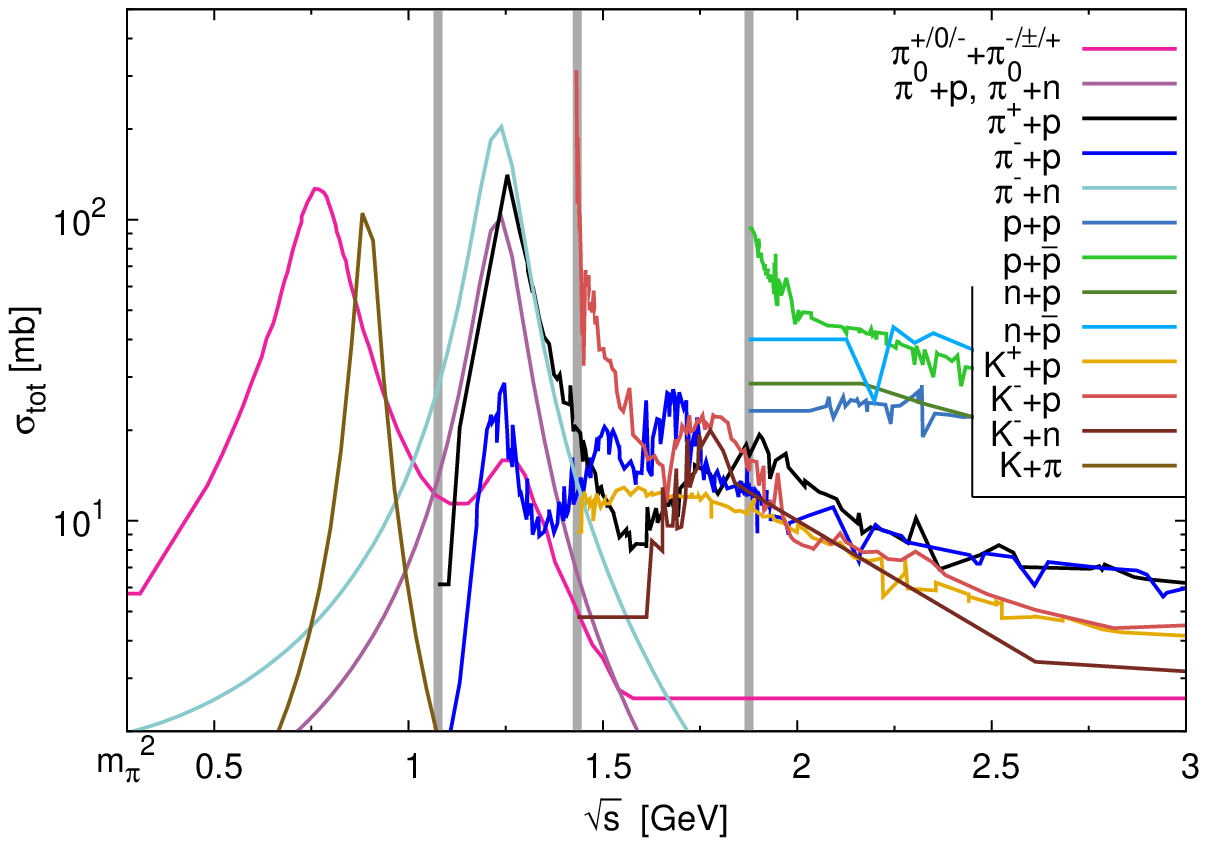}
\caption{Tabulated hadronic cross sections over $\protect\sqrt{s}$ from Ref.~%
\protect\cite{PDG} we used for the Pion-Kaon-Nucleon-Lambda-Sigma gas. The
grey bars denote the minimal $\protect\sqrt{s}$ of the particular scattering
process. The combinations which are not listed here are assumed to be constant \protect\cite{GiBUU,Bass:1998ca,UrQMD2}.}
\label{fig:all_resonances_paper}
\end{figure}

In order to perform these numerical calculations one has to first specify
the differential cross sections for the particle interactions. In this
letter, we restrict ourselves to elastic, isotropic (s-wave) scattering,
employing all available $\sqrt{s}$ dependent cross sections from Ref.~\cite%
{PDG} shown in Fig.~\ref{fig:all_resonances_paper}. Due to the lack of
experimental data, we assume all missing hadronic cross sections to be
constant, as done, e.g. in hadronic transport models~\cite{GiBUU,Bass:1998ca,UrQMD2}. 
The hyperon cross sections thus take constant values between $3-35~\mathrm{mb}$.
We will also make an estimate of the
diffusion coefficients of the QGP. For this purpose, we
assume three flavors of massless quarks and gluons, and choose a unique total
cross section $\sigma _{\mathrm{tot}}$ in such a way that the shear viscosity to entropy density
ratio is fixed to be $\eta /s=1/(4\pi )$, leading to $\sigma _{\mathrm{tot}}\approx 0.72/T^{2}$~\cite{Xu:2007ns,Bouras:2009nn}. Further details on
the choice of the cross sections will be presented in a forthcoming
publication~\cite{FotakisPRD}.

\paragraph*{Results.}
\label{sec:results}

\begin{figure*}[tbp]
\centering
\includegraphics[width=0.9\textwidth]{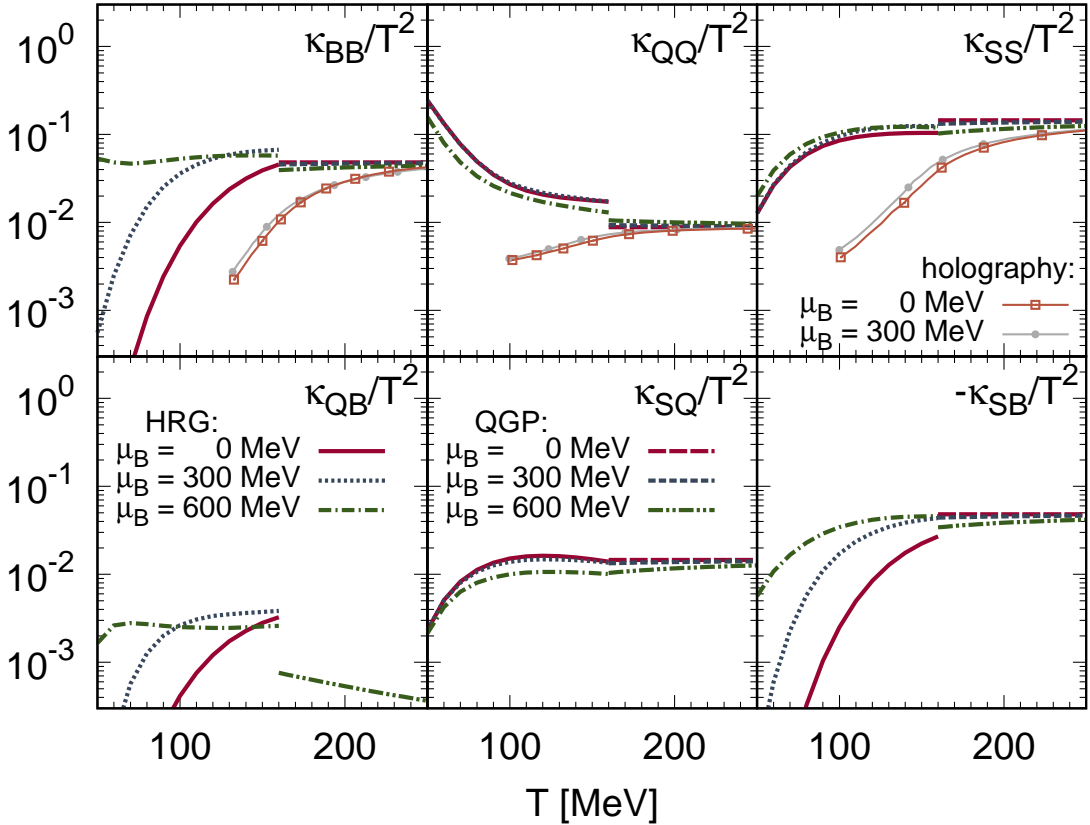}
\caption{All diffusion coefficients for baryon- electric and strangeness
diffusion. The hadronic results include resonance cross sections of the
lightest $19$ hadronic species, whereas the QGP uses massless quarks and
gluons with fixed $4\protect\pi \protect\eta /s=1$. For illustrative
purpose, we show the hadron resonance gas results for $T\leq 160~\mathrm{MeV}
$ and above that the QGP calculation. We compare to holographic results from
Ref.~\protect\cite{Rougemont:2015ona,Rougemont:2017tlu}.}
\label{fig:KappaMatrix_QGP_HG}
\end{figure*}

We first remark that we checked that Onsager's theorem~\cite{PhysRev.37.405,PhysRev.38.2265}, which imposes that $\kappa _{qq^{\prime
}}=\kappa _{q^{\prime }q}$, is fulfilled in all our calculations. We display
our results for the diffusion coefficient matrix from Eq.~\eqref{eq:DiffMatrix} in Fig.~\ref{fig:KappaMatrix_QGP_HG} for $\mu
_{B}=0,300,600~\mathrm{MeV}$. We fix $\mu _{Q}$ and $\mu _{S}$ such that we
always retain an exact Isospin symmetry and vanishing net strangeness, since
this is what approximatly occurs in heavy ion collisions~\cite
{Greiner:1987tg, Cleymans:1992zc}. For illustrative purposes, we show the
HRG results below $T=160~\mathrm{MeV}$, and the QGP results above this
temperature. We also compare here to the non-conformal holographic results
from Ref.~\cite{Rougemont:2015ona,Rougemont:2017tlu}, since these results
are the only ones in literature that contain all three diagonal
coefficients. To the best of our knowledge, the off-diagonal coefficients
have never been calculated before in any model.
 
First we note, that the HRG results are much richer in their $T$ and $\mu_B$ dependence, because of the multitude of scales involved here (masses and resonances). In contrast, the simple choice of a constant $\eta/s$ in the QGP leads to the expected flat behavior for all coefficients~\cite{Arnold:2000dr} (it is known that a running strong coupling lets the coefficients  increase for higher T in the QGP, see, e.g., Refs.~\cite{Greif:2014oia,Aarts:2014nba}). We note that only $\kappa _{QB}$ vanishes at $\mu _{B}=0$ due to the
symmetry of quark charges (and our simplified common cross section). At higher $\mu _{B}$, we see that these coefficients are found to be generally smaller in the QGP phase than they are in hadronic phase ($\kappa _{SQ}$ being the exception). This surprising behavior will be investigated in more detail in a forthcoming paper~\cite{FotakisPRD}.


For the baryon diffusion current $j_{B}^{\mu }$, we expect a strong
dependence on both $\mu _{B}$ and $T$, and indeed this can be seen from the functional behavior of the coefficient $\kappa _{BB}$ in Fig.~\ref{fig:KappaMatrix_QGP_HG}. For $\mu _{B}\lesssim 300~\mathrm{MeV}$ this coefficient rises rapidly with increasing temperature, as the system is less meson dominated at higher temperatures, and mesons act purely as a
resistance for the diffusion of baryons. This effect is also visible in the off-diagonal coefficients $-\kappa_{SB}$ and $\kappa _{QB}$. 
Comparing $\kappa_{QB}$ to $\kappa_{BB}$, in Fig.~\ref{fig:KappaMatrix_QGP_HG}, we infer that the electric charge gradients contribute to the baryon diffusion current about an order of magnitude less than the baryonic gradients. In contrast, gradients in strangeness can be as important as gradients in the baryon charge, as can be seen in the bottom right panel from the magnitude of the coefficient $-\kappa _{SB}$, which is similar in magnitude to $\kappa_{BB}$. We remark that this is due to the hyperons, which carry both $B$
and $S$ charge. The negative sign of $\kappa _{SB}$ indicates that gradients in strangeness
act to reduce the baryon current.

We now discuss the coefficients $\kappa _{QQ},\kappa _{SQ},\kappa _{QB}$, which characterize the
diffusion of electric charges\footnote{
at
$\mu_B=0$, $\kappa_{QQ}/T^2$ is equal to the electric conductivity
$\sigma_{\mathrm{el}}/T$}. We see that $\kappa
_{QQ}/T^{2}$ decreases with temperature, and for increasing values of 
$\mu _{B}$. This happens because the particle density grows, but the ratio
of charged to uncharged species stays the same. The small ratio $\kappa _{QB}/\kappa _{QQ}$ indicates the little importance of baryon chemical potential gradients to the electric diffusion current, whereas $\kappa _{SQ}$ is (for $T\gtrsim 100~\mathrm{MeV}$) of the same order of magnitude as $\kappa _{QQ}$, indicating that strangeness gradients contribute significantly to the electric diffusion current.


Looking at the diffusion coefficients related to strangeness diffusion, we find that $\kappa_{SS}$ is larger than both $-\kappa _{SB}$ and $\kappa _{SQ}$, being even larger in magnitude than the baryon diffusion coefficient (except for very small values of temperature). However, we find that baryonic
gradients act to significantly reduce strangeness currents in both the QGP and HRG, since $\kappa _{SB}$ is negative and its magnitude is only about a factor two smaller than $\kappa _{SS}$. Therefore, it is possible that cancellation effects due to coupling between the currents can lead to small strangeness diffusion currents. On the other hand, $\kappa_{SQ}$ is about an order of magnitude smaller than $\kappa_{SS}$, indicating that electric gradients are less important for strangeness transport. We remark that the $\mu _{B}$ dependence of $\kappa
_{SS},\kappa _{QQ}$ and $\kappa _{SQ}$ is very weak, however their dependence on $\mu _{Q}$ and $\mu _{S}$ can behave differently. This dependence will be addressed in a future publication.

The holographic results from Ref.~\cite{Rougemont:2015ona,Rougemont:2017tlu}
match ours at high $T$ (conformal limit). Their $\mu _{B}$ dependence for
the diagonal coefficients is as weak as for our QGP results, but slightly
lower in magnitude. It is interesting how a simple kinetic calculation, that
simply fixes $\eta /s=1/4\pi $, is already capable of reproducing the basic
trends of such holographic calculations. It would be interesting to see
whether this holds for the off-diagonal coefficients.

\paragraph*{Conclusion.}
\label{sec:conclusion} 
We have calculated the complete diffusion coefficient matrix for the
conserved baryon, electric and strange charges for a hot hadron gas and QGP,
using the traditional Chapman-Enskog formalism. These $6$ transport
coefficients include the baryon diffusion coefficient $\kappa _{BB}$, and the electric and strangeness diffusion coefficients $\kappa _{QQ}$ and $%
\kappa _{SS}$, respectively. We present for
the first time also the three off diagonal transport diffusion coefficients $%
\kappa _{QB},\kappa _{SB}$ and $\kappa _{SQ}$, which describe the mixing
between the different charge currents. In our semi-analytic approach, we
confine ourselves to classical statistics, elastic collisions and isotropic
scattering. We include resonance and measured elastic hadron-hadron
cross sections, when available, taking into account hadrons up to the $\Sigma $
baryons. This constitutes the most extensive result of the charge diffusion matrix in the HRG to date.
For calculations in the QGP phase, we fix $\eta /s$ to be a constant. It is in fact very interesting that most of the diffusion coefficients in the QGP match the HRG results quite well nearby the conjectured phase transition region. 

The diffusion coefficients can be readily used in, e.g., hydrodynamic
simulations, or other model descriptions of high
density heavy ion collisions, where diffusion processes are taken into
account. Those models are and will be increasingly important for low energy
and high density experiments like RHIC BES, NICA or FAIR. Our results
emphasize that the mixing between different diffusion currents is in general important and
should not be neglected when simulating low energy heavy ion collisions.
For example, the contribution to the baryon diffusion current from gradients of
baryon number density can be almost completely canceled by gradients in
strangeness of comparable magnitude, whereas we found electric gradients to be almost negligible for baryon transport.
Electric diffusion is mainly driven by electric and strangeness gradients. 
Strangeness diffusion is mostly affected by strangeness and baryon number gradients, with electric charge gradients being less important. The relevance of these effects for experimental observables remains to be investigated. 
We plan to extend our work to quantum statistics, a more realistic description of the QGP and
possibly more particle species to achieve a fully comprehensive framework of
diffusion properties. 
It would be desirable to compare our results to, e.g.,
lattice QCD results (which at present are only available for the electric conductivity). All coefficients should also be accessible from hadronic transport models, or other dynamical approaches.

\begin{acknowledgments}
\paragraph*{Acknowledgements.}
The authors M.G., J.A.F. and C.G. acknowledge support by the Deutsche
Forschungsgemeinschaft (DFG) through the grant CRC-TR 211 \textquotedblleft
Strong-interaction matter under extreme conditions\textquotedblright . M.G.
acknowledges the support from the \textquotedblleft Helmholtz Graduate School
for Heavy Ion research\textquotedblright. J.A.F. acknowledges support from the  \textquotedblleft Stiftung Polytechnische
Gesellschaft\textquotedblright, Frankfurt am Main. The authors thank Harri Niemi for fruitful discussion.
This work was supported by the Helmholtz International Center for FAIR within the framework of the LOEWE
program launched by the State of Hesse. G.S.D. thanks Conselho Nacional de
Desenvolvimento Cient\'{\i}fico e Tecnol\'{o}gico (CNPq) for financial
support.
\end{acknowledgments}

\bibliographystyle{apsrev4-1}
\bibliography{library_manuell.bib}

\end{document}